\documentclass[12pt,preprint]{aastex}

\usepackage{subfigure}

\newcommand{\goodgap}{\hspace{\subfigtopskip} \hspace{\subfigbottomskip}}

\begin{document}

\title{Study of possible systematics in the $L^*_X$\,-\,$T^*_a$ correlation of Gamma Ray Bursts}

\author{Maria Giovanna Dainotti\altaffilmark{1}, Vincenzo Fabrizio Cardone\altaffilmark{2,3}, Salvatore Capozziello\altaffilmark{3,4}, Michal Ostrowski\altaffilmark{1}, Richard Willingale\altaffilmark{5}}

\altaffiltext{1}{Obserwatorium Astronomiczne, Uniwersytet Jagiello\'nski, ul. Orla 171, 31-501 Krak{\'o}w, Poland, E\,-\,mails\,: mariagiovannadainotti@yahoo.it, mio@oa.uj.edu.pl}

\altaffiltext{2}{Dipartimento di Scienze e Tecnologie dell' Ambiente e del Territorio, Universit\`{a} degli Studi del Molise, \\
Contrada Fonte Lappone, 86090\,-\,Pesche (IS), Italy, E\,-\,mail\,: winnyenodrac@gmail.com}

\altaffiltext{3}{Dipartimento di Scienze Fisiche, Universit\`{a} di Napoli "Federico II", Complesso
Universitario di Monte Sant'Angelo, Edificio N, via Cinthia, 80126 - Napoli, Italy E\,-\,mail\,: capozziello@na.infn.it}

\altaffiltext{4}{I.N.F.N., Sez. di Napoli, Complesso Universitario di Monte
Sant' Angelo, Edificio G, via Cinthia, 80126 - Napoli, Italy}

\altaffiltext{5}{Department of Physics \& Astronomy, University of Leicester, Road Leicester LE1 7RH, United Kingdom, E\,-\,mail\,: rw@star.le.ac.uk}


\begin{abstract}

Gamma Ray Bursts (GRBs) are the most energetic sources in the universe and among the farthest known astrophysical sources. These features make them appealing candidates as standard candles for cosmological applications so that studying the physical mechanisms for the origin of the emission and correlations among their observable properties is an interesting task. We consider here the {\it luminosity $L^*_X$\,-\,break time $T^*_a$} (hereafter LT) correlation and investigate whether there are systematics induced by selection effects or redshift dependent calibration. We perform this analysis both for the full sample of 77 GRBs with known redshift and for the subsample of GRBs having canonical X\,-\,ray light curves, hereafter called $U0095$ sample. We do not find any systematic bias thus confirming the existence of physical GRB subclasses revealed by tight correlations of their afterglow properties. Furthermore, we study the possibility of applying the LT correlation as a redshift estimator both for the full distribution and for the canonical lightcurves. The large uncertainties and the non negligible intrinsic scatter make the results not so encouraging, but there are nevertheless some hints motivating a further analysis with an increased $U0095$ sample.

\end{abstract}


\section{Introduction}

The high fluence values (from $10^{-7}$ to $10^{-5} \ {\rm  erg/cm^2}$) and the huge isotropic energy emitted ($\simeq 10^{50} - 10^{54} {\rm erg}$) at the peak in a remarkably short prompt emission phase make GRBs the most violent and energetic astrophysical phenomena. Fifty years after their discovery in the '60s by the Vela satellites, the nature of GRBs is still unclear. Notwithstanding the variety of their peculiarities, some common features may be identified by looking at their light curves. GRBs have been traditionally classified as {\it short} ($T_{90}<2 s$) and {\it long} ($T_{90}>2s$), although some recent studies (see, e.g., Norris \& Bonnell 2006) have revealed the existence of an intermediate class (IC) thus asking for a revision of this criterium. Consequently, long GRBs have now been divided into two classes, {\it normal} and {\it low luminosity}, the latter one likely being associated with Supernovae. Here, we concentrate our attention on the class of normal long bursts, observed in X\,-\,ray with the aim of better clarifying their origin in the view of possible systematics.
 A valid tool in classifying GRBs is provided by the analysis of their light curves. The data observed by the Beppo\,-\,Sax satellite \cite{Piro01} were reasonably well fitted by a simple phenomenological power\,-\,law expression, $f(t, \nu) \propto t^{-\alpha} \nu^{-\beta}$ with $(\alpha, \beta) \simeq (1.4, 0.9)$. However, a crucial breakthrough in this field has been represented by the launch of the {\it Swift} satellite in 2004. The {\it Swift} instrumental setup, composed by the {\it Burst Alert Telescope} (BAT, $15 - 150$ keV), the {\it X\,-\,Ray Telescope} (XRT, $0.3 - 10$ keV) and the {\it Ultra\,-\,Violet/Optical Telescope} (UVOT, $170 - 650$ nm), allows a rapid follow\,-\,up of the afterglows in different wavelengths giving better coverage of the GRB light curve than the previous missions. Such data revealed the existence of a more complex phenomenology with different slopes and break times thus stressing the inadequacy of a single power\,-\,law function. A significant step forward has been made by the analysis of the X\,-\,ray afterglow curves of the full sample of {\it Swift} GRBs showing that they may be fitted by a single analytical expression \citep{W07} which we referred to in the following as the W07 model.

It is worth stressing that finding out a universal feature would allow us to recognize if GRBs are standard candles looking for correlations among their observables. The $E_{iso}$\,-\,$E_{peak}$ \citep{amati09}, $E_\gamma$\,-\,$E_{peak}$ \citep{Ghirlanda06}, $L$\,-\,$E_{peak}$ \citep{S03} and $L$\,-\,$V$ \citep{R01} correlations are some of the attempts pursued in this direction. However, the problem of large data scatter in the considered luminosity relations \citep{Butler2009,Yu09} and a possible impact of detector thresholds on cosmological standard candles \citep{Shahmoradi09} have been discussed controversially \citep{Cabrera2007}.
Within this wide framework, we consider here the LT correlation between the break time $T_a^* = T_a/(1 + z)$ and the luminosity at the break time $L_X^*$ where $z$ is the GRB redshift and with asteriks we refer to the rest frame quantities. Both these observables refer to the plateau phase of the W07 model. Dainotti et al. (2008) first found that these quantities are not independent, but rather follow the log\,-\,linear relation, $\log L^*_X =  a \log T^*_a + b$, with $a$ and $b$ fixed by the fitting procedure. The LT correlation has then been confirmed \citep{Ghisellini2008,Yamazaki09} and recently updated with 77 GRBs \citep{Dainotti09} leading to the discovery of a new subclass of the afterglows with smooth observed X\,-\,ray light curves, which are prefentially distributed at higher luminosities than the full distribution.

The plan of the paper is as follows. In Section 2, we review the LT correlation explaining how the interested quantities are evaluated and the calibration procedure adopted. Selection effects are discussed in Section 3, while the problem of a possible evolution with $z$ of the calibration parameters is addressed in Section 4. Section 5 investigates the possibility of using the LT correlation as a redshift indicator, while a summary of the results is finally given in Section 6.

\section{The $LT$ correlation}

The LT correlation relates the time scale $T^*_a$ and the X\,-\,ray luminosity $L^*_X$ at $T_a$, where $T_a$ is defined as the end of the plateau phase. Having had the confirmation of the existence of the above correlation \citep{Dainotti09}, we here try to answer the question\,: {\it Is it affected by systematics ?}

As a preliminary remark, let us remember how the quantities of interest are evaluated. The source rest frame luminosity in the {\it Swift} XRT bandpass, $(E_{min}, E_{max})=(0.3,10)$ keV, is computed as\,:

\begin{equation}
L^*_X (E_{min},E_{max},t)= 4 \pi D_L^2(z) \, F_X (E_{min},E_{max},t) \cdot K
\label{eq: lxgeneral}
\end{equation}
where $D_L(z)$ is the GRB luminosity distance at redshift $z$, $F_X$ is the measured X\,-\,ray energy flux (in ${\rm erg/cm^2/s}$) and  $K$ is the K\,-\,correction. Denoting with $f(t)$ the {\it Swift} light curve and following Bloom et al. (2001), we get\,:

\begin{equation}
K F_X(E_{min},E_{max},t) = f(t) \ {\times} \ \frac{\int_{E_{min}/(1 + z)}^{E_{max}/(1 + z)}{E \Phi(E) dE}}{\int_{E_{min}}^{E_{max}}{E \Phi(E) dE}}
\label{eq: fx}
\end{equation}
with $\Phi(E)$ the differential photon spectrum. We model this term as $\Phi(E) \propto E^{-\gamma_a} \propto E^{-(\beta_a + 1)}$ where $(\beta_a, \gamma_a)$ are the spectral and photon index, respectively. It is worth stressing that the fit of the model $\Phi(E)$ is performed considering only the spectrum of the plateau phase, selected using a filter time fixed as $T_a \pm \sigma_{T_a}$; the $T_a$ values together with their errorbars, $\sigma_{T_a}$, are derived in the fitting procedure \citep{W07}. As shown also in previous XRT spectral analysis \citep{Nousek06}, this particular choice of the filter time leads to the single power\,-\,law function as a better fit than the more commonly assumed Band function \cite{Band93}. According to the W07 model, the functional expression for $f(t)$ is\,:

\begin{equation}
f(t) = f_p(t) + f_a(t)
\label{eq: fluxtot}
\end{equation}
where the first term accounts for the prompt (the index "p") $\gamma$\,-\,ray emission and the initial X\,-\,ray decay, while the second one describes the afterglow (the index "a"). Both components are modeled with the same functional form\,:

\begin{equation}
f(t) = \left \{
\begin{array}{ll}
\displaystyle{F_c \exp{\left ( \alpha_c - \frac{t \alpha_c}{T_c} \right )} \exp{\left (
- \frac{t_c}{t} \right )}} & {\rm for} \ \ t < T_c \\
~ & ~ \\
\displaystyle{F_c \left ( \frac{t}{T_c} \right )^{-\alpha_c}
\exp{\left ( - \frac{t_c}{t} \right )}} & {\rm for} \ \ t \ge T_c \\
\end{array}
\right .
\label{eq: fc}
\end{equation}
where $c$ = $p$ or $a$. The transition from the exponential to the power law occurs at the point $(T_{c},F_{c})$ where the two functional sections have the same value and gradient. The parameter $\alpha_{c}$ is the temporal power law decay index  and the time $t_{c}$ is the the initial rise time scale.  We refer to Willingale et al. (2007) for further details on the analysis, while we only remind here that a usual $\chi^2$ fitting of the $\log{(flux)}$ vs $\log{(time)}$ data provides estimates and uncertainties on the time parameters $(\log{T_p}, \log{T_a})$ and the products $(\log{F_p T_p}, \log{F_a T_a})$.

For the afterglow part of the light curve, we have computed values $L^*_X$ (eq. \ref{eq: lx}) at the time $T_a$, which marks the end of the plateau phase and the beginning of the last power law decay phase. We have considered the following approximation which takes into accounts the functional form, $f_a$, of the afterglow component only:

\begin{equation}
f(T_a) \approx f_a(T_a)=F_a \exp{\left ( - \frac{T_p}{T_a} \right )}
\label{eq: fluxafterglow}
\end{equation}
where we set $t_a = T_p$ because in most cases the afterglow component is fixed at the transition time of the prompt emission, $T_p$. Actually, we are using Eq.(\ref{eq: fluxafterglow}), instead of  (\ref{eq: fluxtot}) since the contribution of the prompt component is typically smaller than $5\%$, much lower than the statistical uncertainty on $f_a(T_a)$. Neglecting $f_p(T_a)$ thus allows to reduce the error on $F_X(T_a)$ without introducing any bias. This latter error is then estimated by simply propagating those on $\beta_a$, $\log{T_a}$ and $\log{F_a T_a}$ thus implicitly assuming that their covariance is null. Inserting Eqs.(\ref{eq: fluxafterglow}) and (\ref{eq: fx}) into Eq.(\ref{eq: lxgeneral}), one then obtains\,:

\begin{equation}
L^*_X = \frac{4\pi D_L^2(z) F_X }{(1+z)^{1-\beta_{a}}}
\label{eq: lx}
\end{equation}
where $F_X = F_a \exp{(-T_p/T_a)}$ is the observed flux at the time $T_a$.

 As a final important remark, we note that the presence of the luminosity distance $D_L(z)$ in Eq.(\ref{eq: lx}) constrains us to adopt a cosmological model to compute $L^*_X$. We then use a flat $\Lambda$CDM model so that the luminosity distance reads\,:

\begin{equation}
D_L(z) = \frac{c}{H_0} (1 + z) \int_{0}^{z}{\frac{dz'}{\sqrt{\Omega_M (1 + z')^3 +
(1 - \Omega_M)}}} \ .
\label{eq: dl}
\end{equation}
In agreement with the WMAP seven year results \cite{WMAP7}, we set $(\Omega_M, h) = (0.272, 0.704)$ with $h$ the Hubble constant $H_0$ in units of $100 \ {\rm km/s/Mpc}$.

\subsection{Calibration parameters}

Let us suppose that $R$ and $Q$ are two quantities related by a linear relation

\begin{equation}
R = a  Q + b
\end{equation}
and denote with $\sigma_{int}$ the intrinsic scatter around this relation. Calibrating such a relation means determining the two coefficients $(a, b)$ and the intrinsic scatter $\sigma_{int}$. To this aim, we will resort to a Bayesian motivated technique \cite{Dago05} thus maximizing the likelihood function ${\cal{L}}(a, b, \sigma_{int}) = \exp{[-L(a, b, \sigma_{int})]}$ with\,:

\begin{equation}
L(a, b, \sigma_{int}) = \frac{1}{2} \sum{\ln{(\sigma_{int}^2 + \sigma_{R_i}^2 + a^2 \sigma_{Q_i}^2)}} +
\frac{1}{2} \sum{\frac{(R_i - a Q_i - b)^2}{\sigma_{int}^2 + \sigma_{Q_i}^2 + a^2 \sigma_{Q_i}^2}}
\label{eq: deflike}
\end{equation}
where the sum is over the ${\cal{N}}$ objects in the sample. Note that, actually, this maximization is performed in the two parameter space $(a,
\sigma_{int})$ since $b$ may be estimated analytically as\,:

\begin{equation}
b = \left [ \sum{\frac{R_i - a Q_i}{\sigma_{int}^2 + \sigma_{R_i}^2 + a^2
\sigma_{Q_i}^2}} \right ] \left [\sum{\frac{1}{\sigma_{int}^2 + \sigma_{R_i}^2 + a^2
\sigma_{Q_i}^2}} \right ]^{-1}
\label{eq: calca}
\end{equation}
so that we will not consider it anymore as a fit parameter. The above formulae easily applies to our case setting $R = \log L^*_X(T_a)$ and $Q = \log T^*_a$. We estimate the uncertainty on $\log L^*_X(T_a)$ by propagating the errors on $(T_a, T_p, F_a T_a, \beta_a)$.

The Bayesian approach used here also allows us to quantify the uncertainties on the fit parameters. To this aim, for a given parameter $p_i$, we first compute the marginalized likelihood ${\cal{L}}_i(p_i)$ by integrating over the other parameter. The median value for the parameter
$p_i$ is then found by solving\,:

\begin{equation}
\int_{p_{i,min}}^{p_{i,med}}{{\cal{L}}_i(p_i) dp_i} = \frac{1}{2}
\int_{p_{i,min}}^{p_{i,max}}{{\cal{L}}_i(p_i) dp_i} \ .
\label{eq: defmaxlike}
\end{equation}
The $68\%$ ($95\%$) confidence range $(p_{i,l}, p_{i,h})$ are then found by solving\,:

\begin{equation}
\int_{p_{i,l}}^{p_{i,med}}{{\cal{L}}_i(p_i) dp_i} = \frac{1 - \varepsilon}{2}
\int_{p_{i,min}}^{p_{i,max}}{{\cal{L}}_i(p_i) dp_i} \ ,
\label{eq: defpil}
\end{equation}

\begin{equation}
\int_{p_{i,med}}^{p_{i,h}}{{\cal{L}}_i(p_i) dp_i} = \frac{1 - \varepsilon}{2}
\int_{p_{i,min}}^{p_{i,max}}{{\cal{L}}_i(p_i) dp_i} \ ,
\label{eq: defpih}
\end{equation}
with $\varepsilon = 0.68$ (0.95) for the $68\%$ ($95\%$) range respectively.

\section{Threshold selection of the fit error parameter}

Dainotti et al. (2010, hereafter D10) have recently updated the LT correlation using a sample of 77 GRBs with known redshift and {\it Swift} X\,-\,ray afterglow light curves. D10 have defined a fit error parameter $u \equiv \sqrt{\sigma_{L^*_{X}}^2 + \sigma_{T^*_a}^2}$, measured in the burst rest frame, to analyze how accuracy of fitting the canonical lightcurve, (eq. \ref{eq: fluxtot} and \ref{eq: fc}) to the data influences the studied correlations. This definition is used to distinguish the canonical shaped light curves from the more irregular ones, perturbed by secondary flares and various non uniformities. D10 have then defined a fiducial sample selecting only GRBs with $u < 4$ and excluding the IC objects thus selecting 62 out of the original 77 GRBs. To be consistent with D10, we here still consider only the fiducial sample. As a general remark, we would like to stress that the study of a whatever correlation among GRBs observables should involve only physically homogenous subsamples thus motivating our exclusion of the IC GRBs because of their different properties from the long ones that mainly constitute our sample. In other words, with homogenous sample we indicate a subsample of GRBs that have lightcurves well defined, in the sense that the Willingale model represents with very good accuracy the parameters values representing the afterglow and the plateau.
As a consequence this subsample tightly obeys to the correlation and from this evidence we infer that the properties of the GRBs in this subset are the same. For example, the XRFs are included in the subsample, since they obey to the correlations, giving in this way evidence to the theory according to which they have the same progenitor mechanism of the normal long GRBs.

\begin{figure*}[t]
\centering
\includegraphics[width=10cm,angle=-90]{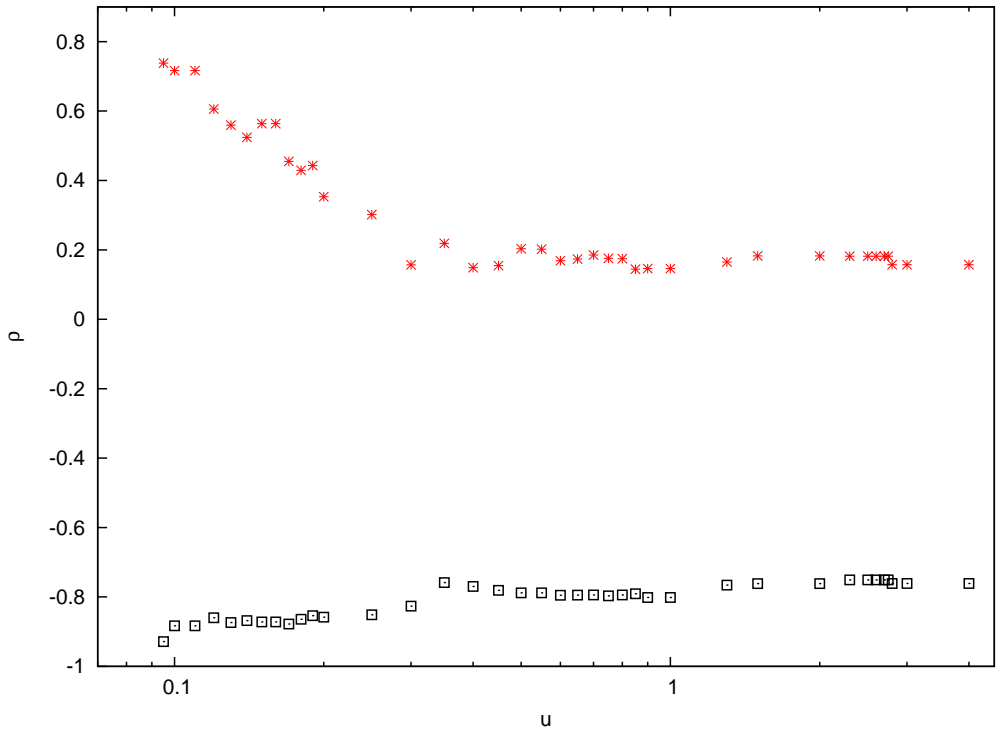}
\caption{Spearman correlation coefficients for the $L_X^*$\,-\,$T_a^*$ (black squares) and $\beta_a$\,-\,$T_a^*$ (red asterisks) as function of the threshold $u$ value.}
\label{rho}
\end{figure*}

\begin{figure*}[t]
\centering
\subfigure{\includegraphics[width=6cm]{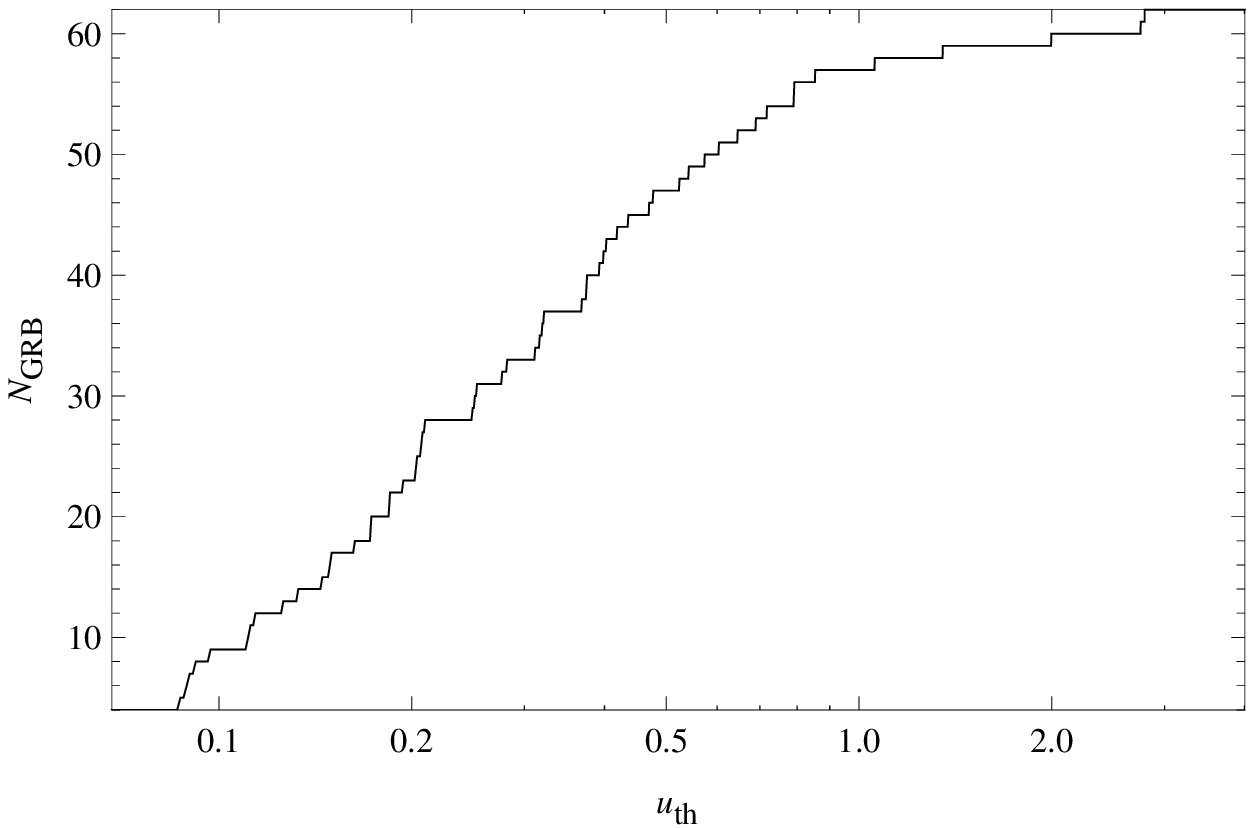}} \goodgap
\subfigure{\includegraphics[width=6cm]{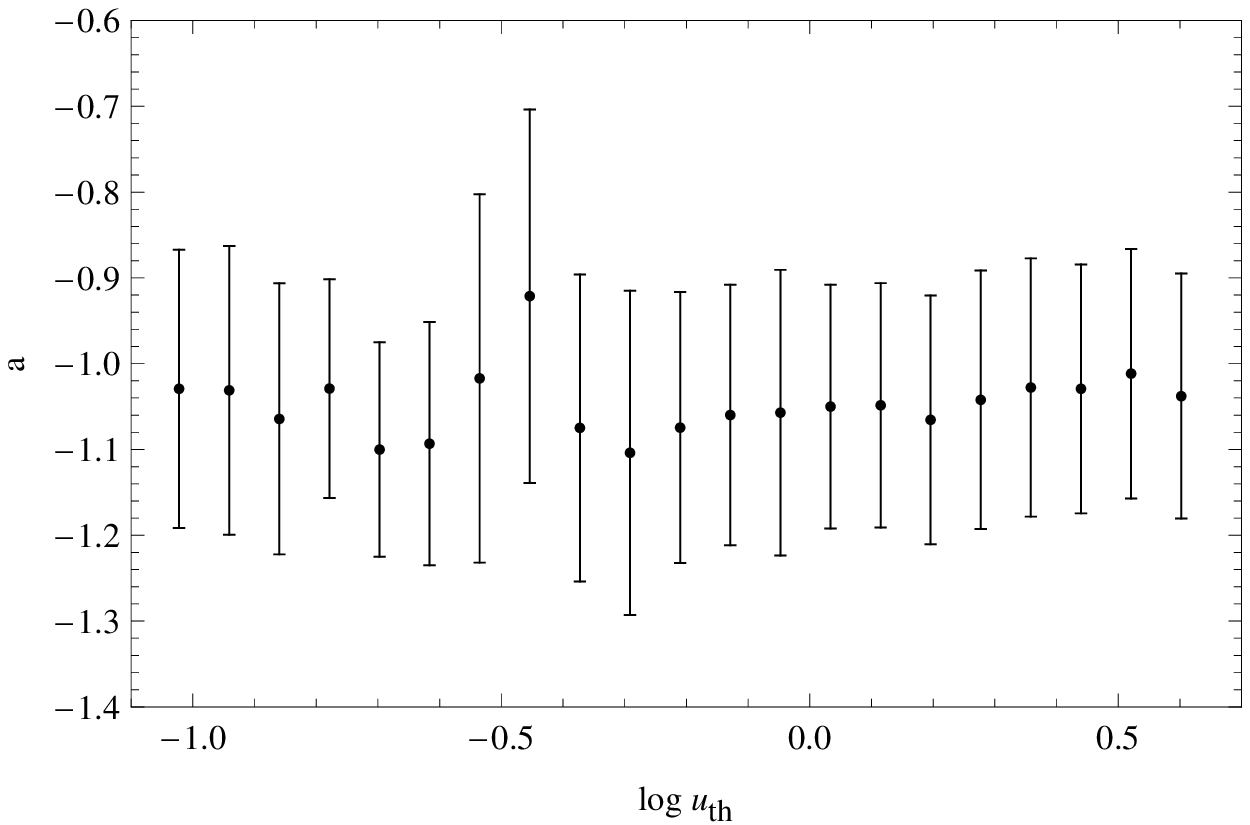}} \goodgap \\
\subfigure{\includegraphics[width=6cm]{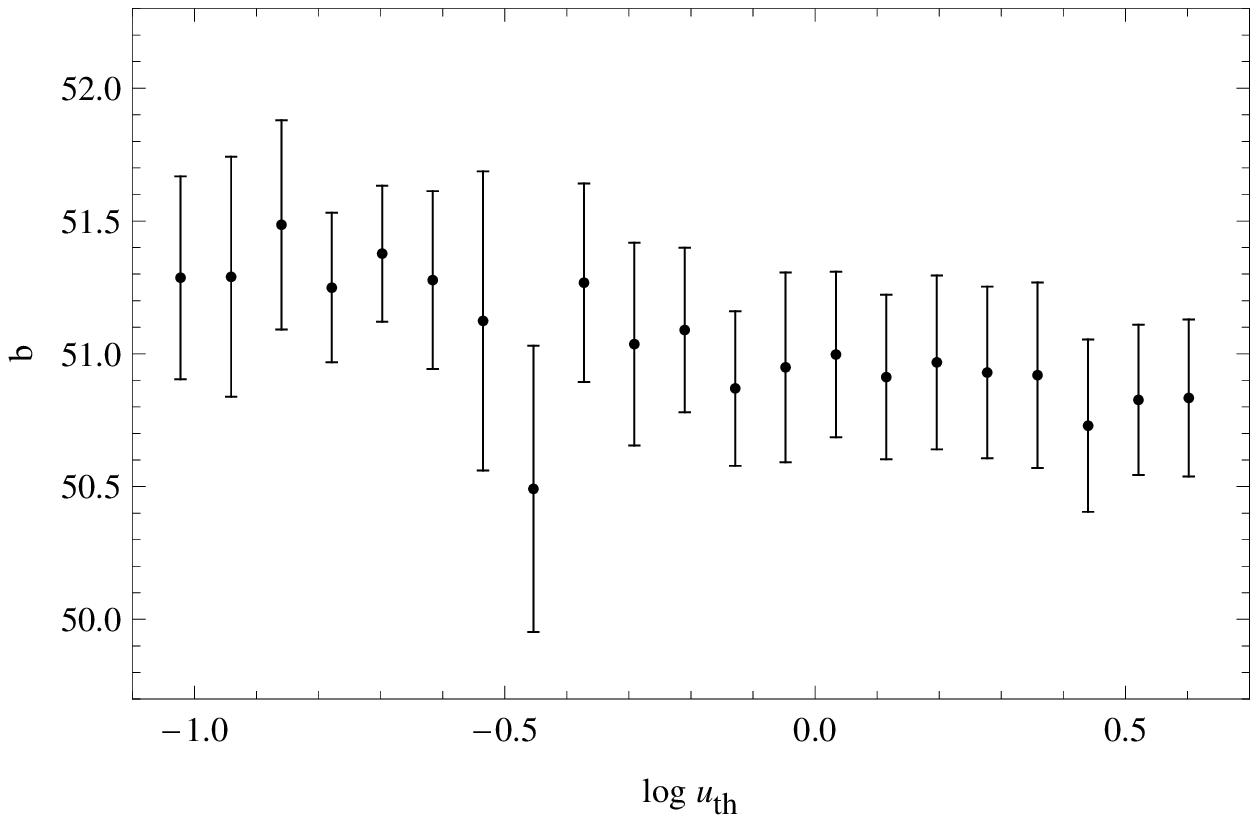}} \goodgap
\subfigure{\includegraphics[width=6cm]{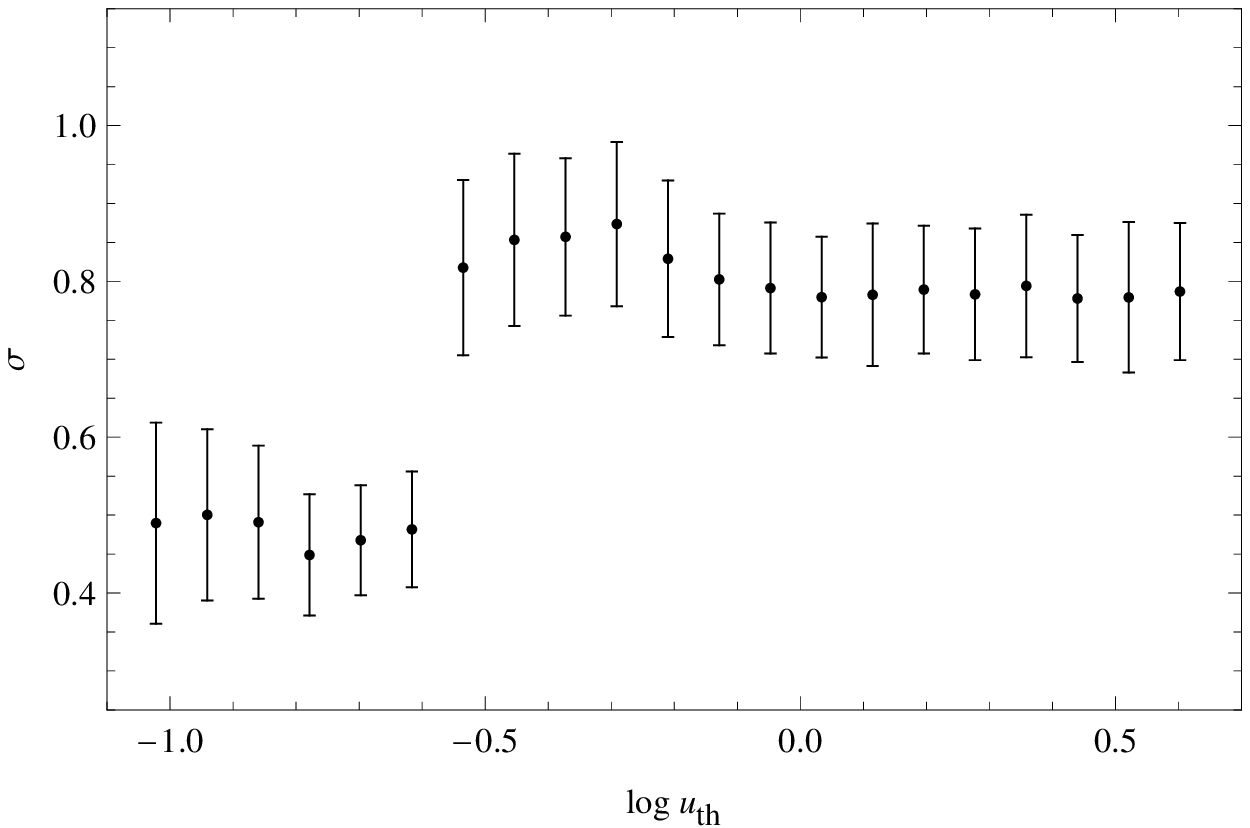}} \goodgap \\
\caption{Number of GRBs, calibration parameters $(a, b)$ and intrinsic scatter $\sigma_{int}$ as function of the threshold $u_{th}$ value. See text for a detailed discussion.}
\label{fig: bestfitsn}
\end{figure*}

As a first indicator for the existence of a relation, we use the Spearman correlation coefficient $\rho$ \cite{Spearman} providing a non\,-\,parametric measure of the statistical significance of the dependence between two quantities. Fig.\,\ref{rho} shows $\rho_{LT} = \rho(L_X^*, T_a^*)$ as a function of the threshold $u_{th}$ value, used to exclude from the fiducial sample GRBs with $u > u_{th}$. As we can note, the smaller $u_{th}$ is, the larger $\rho$ is, i.e. the more we are confident that a statistically meaningful correlation indeed exists. The same figure also shows a similar analysis for $\rho_{\beta T} = \rho(\beta_a, T_a^*)$ suggesting that also the slope $\beta_a$ of the GRB spectrum and the break time $T_a^*$ are correlated. It is worth noting that the smaller $u$, the smaller the error on $(\log{T_a^*}, \log{L_X^*})$ is. Examining the light curves, we find that small errors are obtained for the GRBs that follow better the W07 model. We therefore argue that both the LT and $\beta_a$\,-\,$T_a^*$ correlations are statistically meaningful provided the GRBs in the sample belong to the class well described by the W07 model.

In order to better investigate the impact of the $u$ selection, we fit the LT correlation to GRBs subsamples obtained by selecting only those objects with $u < u_{th}$ with $u_{th}$ running from 0.095 to 4 in steps of 0.01 (with $u_{th} = 4$ for the fiducial sample and $u_{th} = 0.095$ for the canonical ones). The upper left panel in Fig.\,\ref{fig: bestfitsn} shows that the number of GRBs in the sample obviously increases with $u_{th}$, but the price to pay is including GRBs with large errors on $(\log{T_a^*}, \log{L_X^*})$. Such large uncertainties may be due to bad sampling or to a less precise determination of the parameters fitted within the W07 model. In both cases, the estimated values of the fit parameters $(\log{T_p}, \log{T_a}, \log{F_a T_a})$ are not reliable so that it is a safer option to not include large $u$ GRBs in the analysis of correlations. Our chosen value $u = 4$ represents a compromise between the need to assemble a statistically meaningful sample and avoiding uncertain couples $(\log{T_a^*}, \log{L_X^*})$ that can unnecessarily increase the intrinsic scatter.

As a first interesting result, we find that the intrinsic scatter $\sigma_{int}$ is smaller for smaller $u_{th}$ selected samples, with a sharp drop for $u_{th} = 0.4$. The high values of $\rho_{LT}$ and the decreasing scatter point towards a scenario where the GRBs most deviating from the LT correlation are actually the ones with the less precise determination of the fitted parameters consistent with our guess that their estimated values of $\log T^*_a$ and $\log L^*_X$ are not reliable.

\begin{figure*}[th]
\includegraphics[width=15cm]{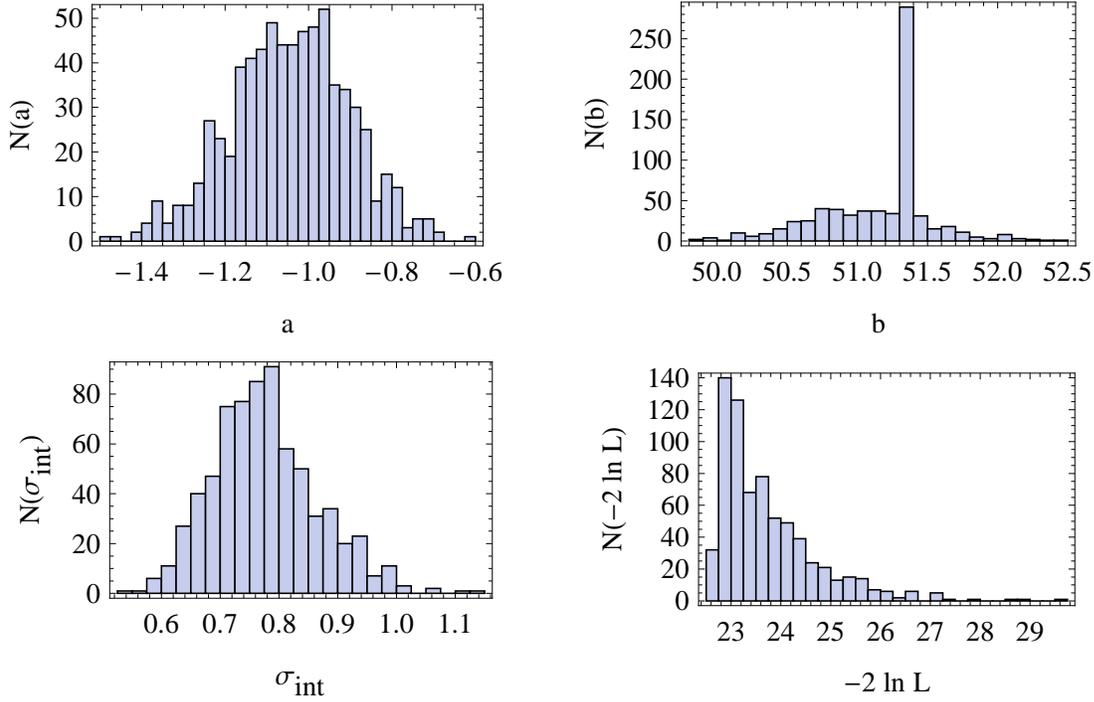}
\caption{Distributions of $(a, b, \sigma_{int})$ and $-\ln{L}$, defined by Eq.(\ref{eq: deflike}), for the fiducial sample (62 GRBs with $u < 4$).}
\label{probability}
\end{figure*}

\begin{figure*}[th]
\includegraphics[width=15cm]{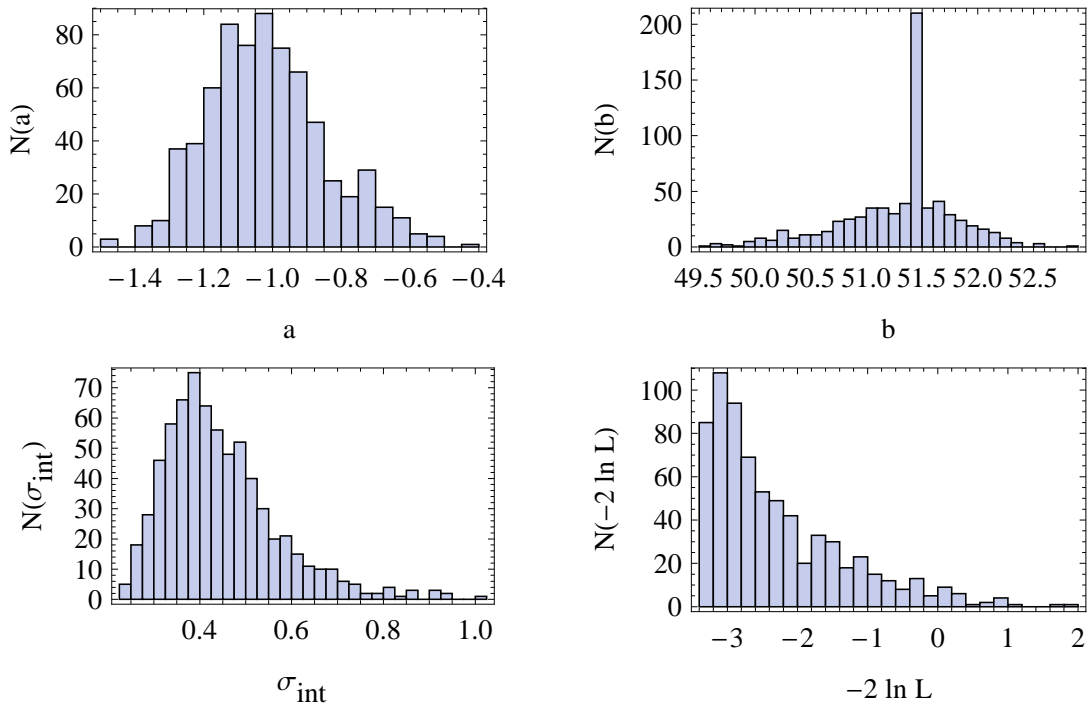}
\caption{Same as Fig.\,\ref{probability} but for the UP sample (8 GRBs with $u < 0.095$).}
\label{probabilityUP}
\end{figure*}

\begin{figure*}
\centering
\subfigure{\includegraphics[width=6cm]{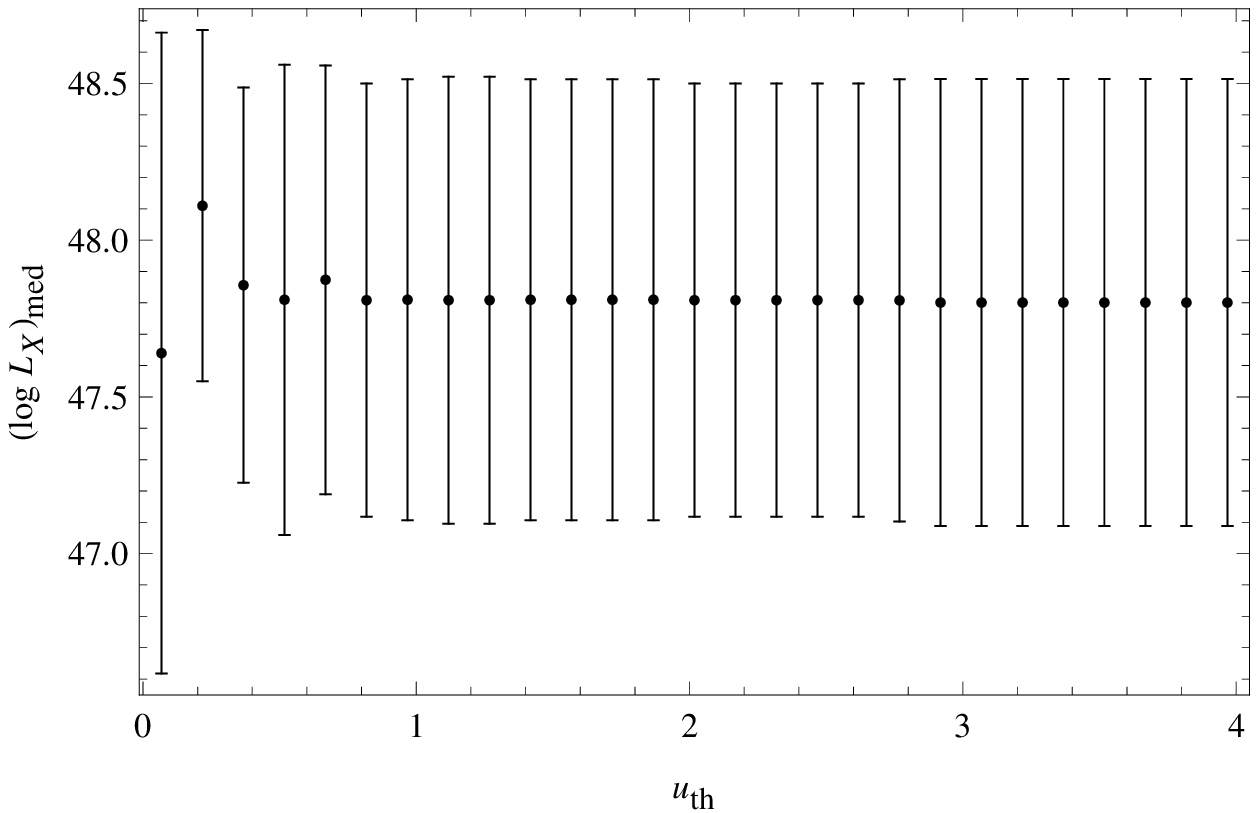}} \goodgap
\subfigure{\includegraphics[width=6cm]{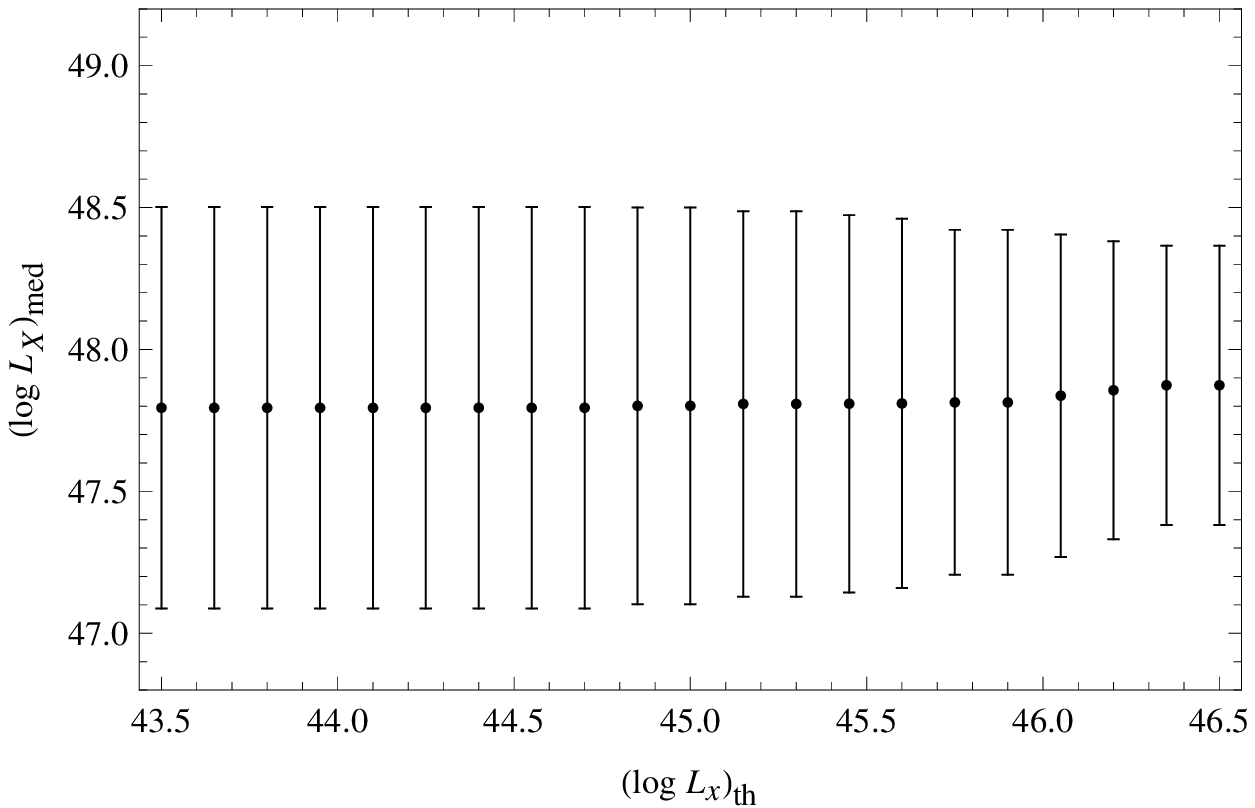}} \goodgap \\
\subfigure{\includegraphics[width=6cm]{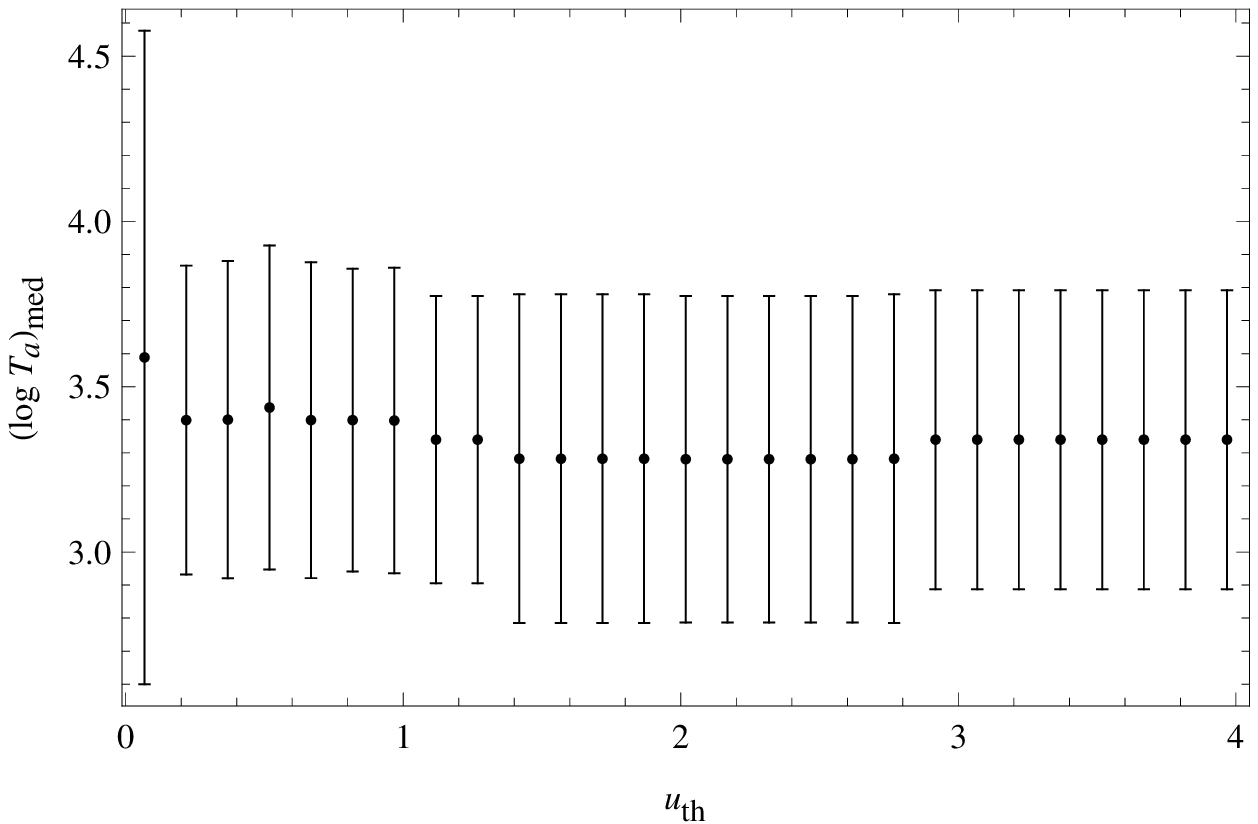}} \goodgap
\subfigure{\includegraphics[width=6cm]{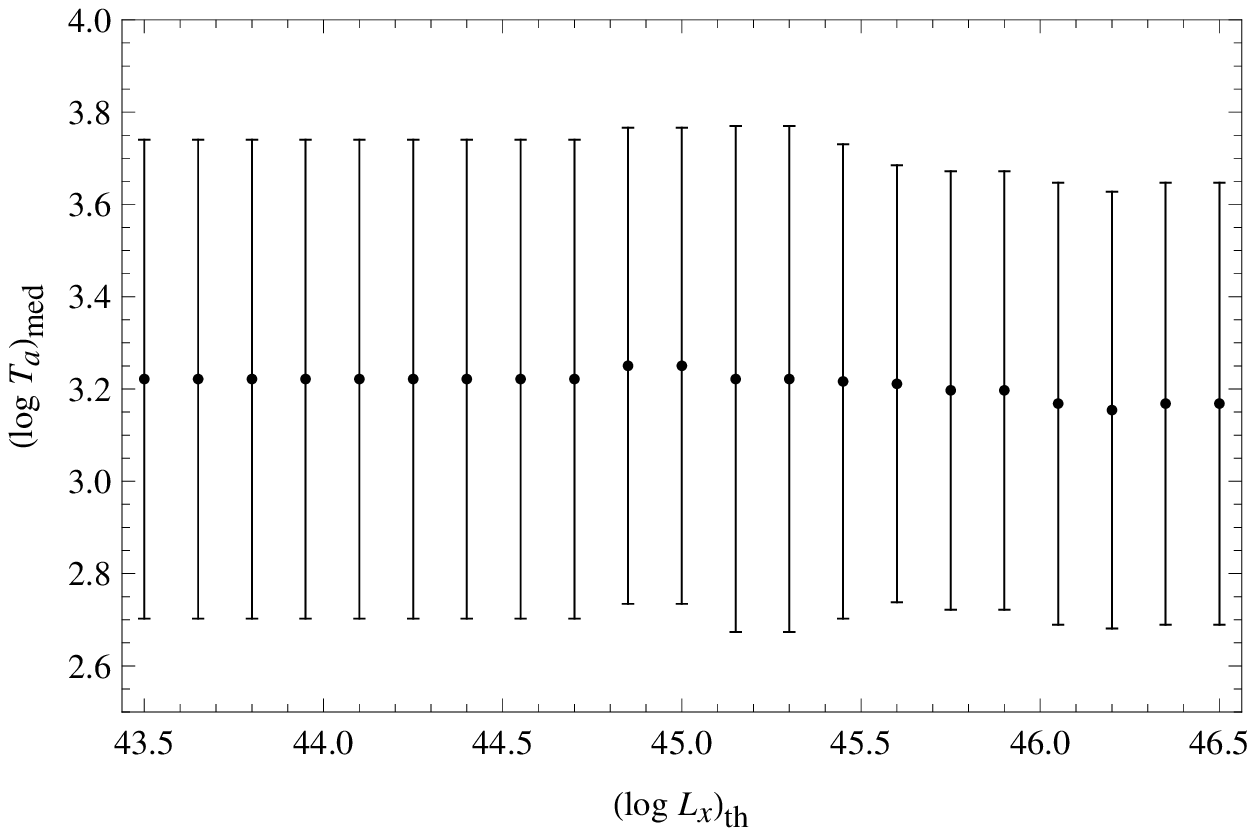}} \goodgap \\
\subfigure{\includegraphics[width=6cm]{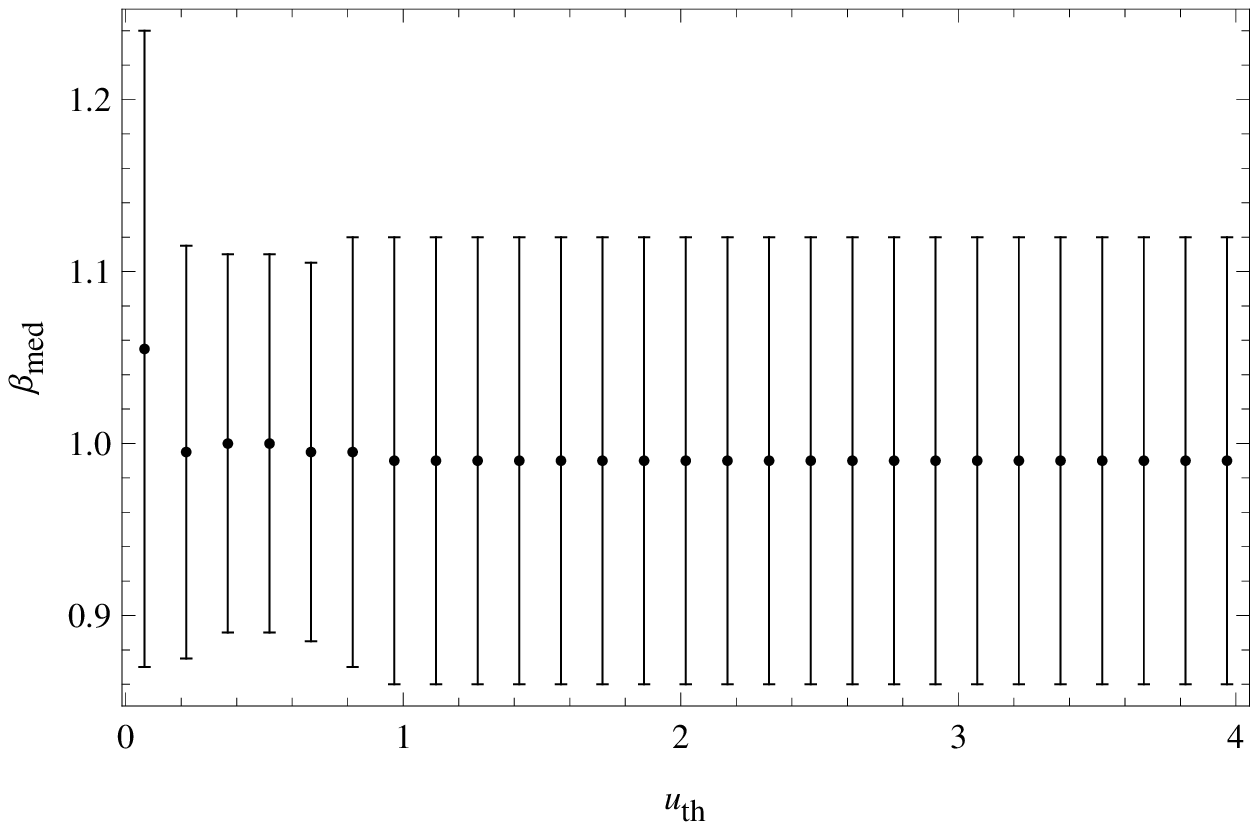}} \goodgap
\subfigure{\includegraphics[width=6cm]{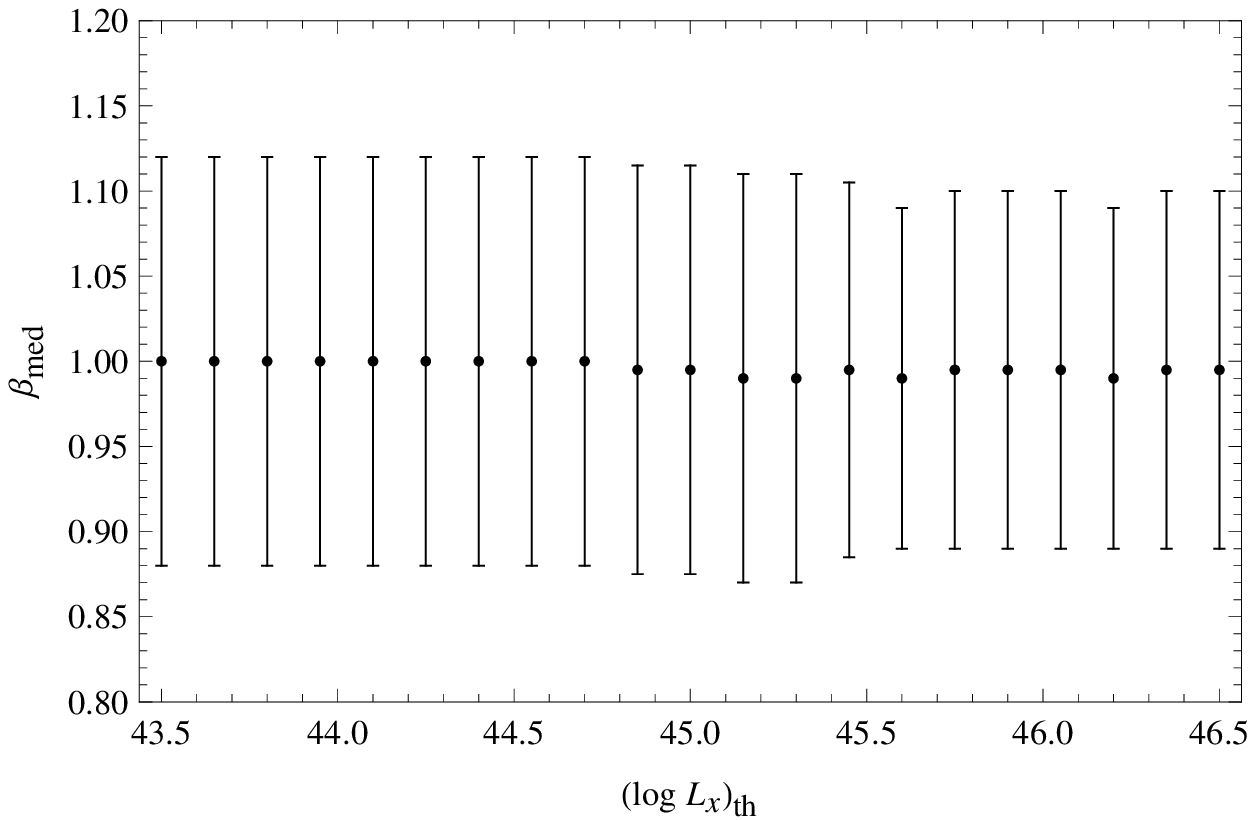}} \goodgap \\
\caption{Median and median deviation values of $(\log{L_X^*}, \log{T_a^*}, \beta_a)$ as function of the threshold values of $u$ (left panels) and $\log{L_X^*}$ (right panel).}
\label{fig: uthlxtest}
\end{figure*}

It is worth wondering whether selecting on $u$ biases the calibration parameters $(a, b)$. Upper right and lower left panels in Fig.\,\ref{fig: bestfitsn} indeed shows a clear increase of $b$ as $u_{th}$ gets smaller, while the slope $a$ of the correlation remains almost constant at the value $a \simeq -1.06$, consistent with the results in D10 that the small $u$ GRBs (referred to as canonical GRBs in D10) define an subsample $U0095$ for the LT correlation. Actually, one has also to consider the error bars on the fitted parameters although we remember that $b$ is actually correlated to $a$ and $\sigma_{int}$ being analitically set by Eq.(\ref{eq: calca}). When the large error bars are taken into account, $a$ can indeed be considered independent on $u_{th}$, while the trend with $u_{th}$ of the zeropoint $b$ remains meaningful.  We place here a general explanation on the size of the error bars presented in Fig. \ref{fig: bestfitsn}, Fig. \ref {fig: uthlxtest}, Fig. \ref{fig: deltamu} and Fig. \ref{fig: muvsz}, namely how the uncertainties on the calibration parameters have been derived. The size of the uncertainties does not reflect only the scatter in the data in the plots, in fact we can note that they are greater than $ 1 \sigma$, since they reflect also the intrinsic scatter in the law $\log L^*_X =  a \log T^*_a + b$. Furthermore, in Fig. \ref {fig: uthlxtest} the errorbars represented are not directly obtained from the Equ. \ref{eq: lx} but they are computed as the median absolute deviation from the median of the parameters. The median values of the observables in the GRBs lightcurves present highly scatter since they reflect intrinsic inhomogeneities in the parameters values. As discussed in detail in our previous papers \citep{DCC,Dainotti09}, the constraints on $(a, b, \sigma_{int})$ have been obtained by running a Monte Carlo Markov Chain algorithm to explore the parameter space $(a, \sigma_{int})$, while $b$ is analytically derived through Eq.(\ref{eq: calca}). To this end, we run two chains, check convergence through the Gellman\,-\,Rubin test \citep{Gelman1992} and finally merge them to estimate the median value and the $68\%$ and $95\%$ confidence ranges. Figs.\,\ref{probability} and \ref{probabilityUP} show these histograms \footnote{A caveat is related to the high peak in the $b$ histogram. Because of parameters degeneracy, there will be different possibilities to get a value of $b$ close to the best fit one. Since the code preferentially selects models with $(a, \sigma_{int})$ as close as possible to the best fit parameters, we will get a lot of couples $(a, \sigma_{int})$ giving almost the same $b$ value. In order to show the full $b$ distribution, we have chosen a range much larger than what is actually needed so that the central bin (i.e., the one which the best fit $b$ lies within) will be much more populated than the other ones hence explaining the peak in the figures. Note also that the first bin in the $P(-2 \ln{L})$ plots is less populated because it is the one corresponding to the best fit parameters. In order to reach convergence, the MCMC code must first find the best fit and then move away from here so that the first bin is not the most populated one.} for the fits to the 77 GRBs with $u < 4$ and the U0095 sample. Therefore, the error bars in Fig.\,\ref{fig: bestfitsn} are determined not only by the scatter in the data, but also by the degeneracies among the model parameters. Moreover, being the distributions mildly asymmetric, the $68\%$ confidence range should not be meant as $1\sigma$ error although we use these abuse of terminology for sake of simplicity. When comparing the results of the fits to the different $u$ selected samples, one should therefore compare the histograms on $(a, b, \sigma_{int})$ and then consider the calibration parameters of different fits in agreement if the corresponding histograms well overlap. This is, for instance, the case for the fiducial and U0095 samples. We note that the median values of the distributions are different, being ($a, b, \sigma_{int}) = (-1.04, 51.30, 0.76$) for the fiducial sample and ($a, b, \sigma_{int}) = (-1.05, 51.40, 0.40$) for the U0095 one. However, the histograms for $a$ well overlap so that we find no statistical meaningful difference, while this is the case for both $b$ (although weak) and $\sigma_{int}$. The error bars plotted in Fig.\,\ref{fig: bestfitsn} allow a quick check for the samples with varying $u$ making us confident that the trends commented above based only on the median values are statistically meaningful.

Up to now, we have interpreted the selection on $u$ as a way to find out the GRBs most closely following the W07 model. It is worth wondering whether such a selection biases in some way the sample by selecting only, e.g., high luminosity GRBs or the shortest ones. To this end, we show in Fig.\,\ref{fig: uthlxtest} the median values (with the median deviation) of $(\log{L_X^*}, \log{T_a^*}, \beta_a)$ as function of the threshold value used for the error parameter $u$.  As these plots clearly show, there is no trend of the median values of $(\log{L_X^{\star}}, \log{T_a^{\star}}, \beta_a)$ with $u$, namely the invariance of $u$ with respect to $(\log{L_X^*}, \log{T_a^*})$ corresponds directly to invariance in the samples that result when GRBs are chosen for certain $u$. Indeed, even neglecting the large error bars \footnote{Note that we are here using the median deviation to characterize the width of the distribution so that, strictly speaking, these are not $1 \sigma$ errors. Note also that the first point in every distribution has a larger error bar since the corresponding sample is made out of only 4 GRBs with $u \le u_{th}$ and $u_{th} < 0.095$. We still plot this point for completeness although it is likely statistically meaningless.}, the median values keep constant showing that the samples selected by imposing $u \le u_{th}$ sample the same region in the parameter space $(\log{L_X^{\star}}, \log{T_a^{\star}}, \beta_a)$. This is due to the fact that the definition of $u$ depends directly on $\sigma_{L^*_{X}}^2$ and $\sigma_{T^*_a}^2$. Therefore, the direct dependence on the possible biases on the sample depends on the parameter values that characterize $u$. Nevertheless, to be confident that correlation among the parameters will not affect the sample selection we have tested also the median values of $\beta_a$ vs $u$, because there is an indication of a correlation among $\log T_a$ vs $\beta_a$ especially for the limiting $u=0.095$ sample, \citep{Dainotti09}. All the tests described  clearly shows that the selection on $u$ is only a way to find out the GRBs following as close as possible the Willingale's model, but this criterion don't bias the samples. As a consequence, we can conclude that the smaller scatter of the LT correlation for canonical GRBs is not a product of selection effects, but rather the outcome of a (still to be understood) physical mechanism.

A careful inspection of the $\log{L_X^*}$ vs $\log{T_a^*}$ plot suggests that the most deviating points are the low luminosity GRBs. We have therefore repeated the above analysis by selecting samples with $\log{L_X^*} > (\log{L_X^*})_{th}$ with $(\log{L_X^*})_{th}$ running from $43.50$ to $46.50$ in steps of 0.15. As shown in the right panels of Fig.\,\ref{fig: uthlxtest}, such a selection criterion do not bias the sample in $(\log{L_X^*}, \log{T_a^*}, \beta_a)$ hence suggesting that the intrinsic scatter of the LT correlation could be reduced by using only moderately bright GBRs. However, since we do not have a physical motivation for applying such a criterion, we have not carried out this analysis.

\begin{table}[t]
\caption{Results of the calibration procedure for GRBs divided in three equally populated redshift bins with $(z_{min}, z_{max}) = (0.08, 1.56)$,
$(1.71, 3.08)$, $(3.21, 8.26)$ for bins Z1, Z2, Z3.}\label{Table1}
\begin{center}
\begin{tabular}{cccccc}
\hline
Id & $\rho_{LT}$ & $(a_{bf}, b_{bf}, (\sigma_{int})_{bf})$ & $a_{median}$ & $b_{median}$ & $(\sigma_{int})_{median}$ \\
\hline
$~$ & $~$ & $~$ & $~$ & $~$ & $~$ \\
Z1 & -0.69 & (-1.20, 51.04, 0.98) & $-1.08_{-0.30}^{+0.27}$ & $51.05_{-0.33}^{+1.27}$ & $1.01_{-0.16}^{+0.20}$\\

$~$ & $~$ & $~$ & $~$ & $~$ & $~$ \\

Z2 & -0.83 & (-0.90, 50.82, 0.43) & $-0.86_{-0.16}^{+0.18}$ & $50.90_{-0.70}^{+0.27}$ & $0.45_{-0.08}^{+0.09}$ \\
$~$ & $~$ & $~$ & $~$ & $~$ & $~$ \\

Z3 & -0.63 & (-0.61, 50.14, 0.26) & $-0.58_{-0.15}^{+0.14}$ & $50.15_{-0.49}^{+0.25}$ & $0.26_{-0.06}^{+0.07}$\\
$~$ & $~$ & $~$ & $~$ & $~$ & $~$ \\

\hline
\end{tabular}
\end{center}
\end{table}

\section{A redshift dependent calibration ?}

The redshift range covered by our GRBs sample is quite large with $(z_{min}, z_{max}) = (0.08, 8.26)$ although the distribution is actually quite inhomogenous. Indeed, we have a single GRB at $z = 8.26$ with the second farthest GRBs being at $z = 5.3$. Similarly, the closest GRBs is at $z = 0.08$, but the second one is at $z = 0.12$. The redshift distribution has played, up to now, no role in our analysis since the LT correlation has been fitted to the full GRBs sample thus implicitly assuming that the calibration coefficients $(a, b, \sigma_{int})$ are the same over this wide redshift range. It is worth wondering whether this is actually the case. To this end, we have therefore recalibrated the $LT$ correlation dividing the GRBs in three equally populated redshift bins. Note that we use here the fiducial sample ($u < 4$) in order to have a good statistics. If we had chosen, for instance, a set with $u <0.3$, only 11 GRBs per bin would be present thus leading to large errors preventing any comparison.

\begin{figure*}
\centering
\includegraphics[width=14cm]{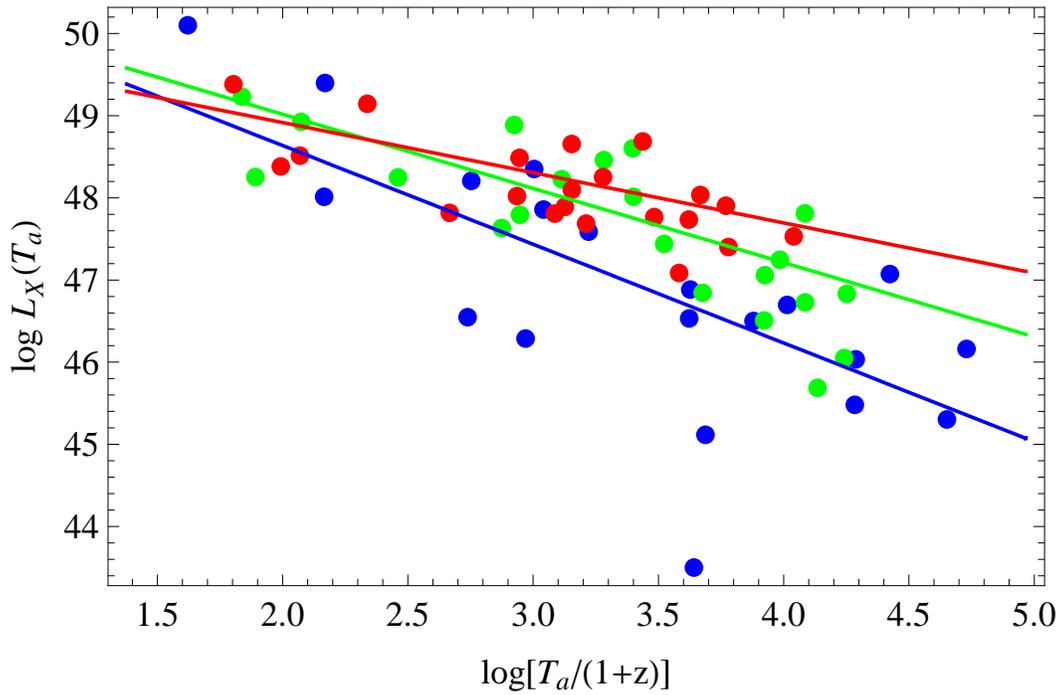}
\caption{$\log{L_X^*}, \log{T_a^*}$ correlation divided in the three redshift bins $Z1=(0.08, 1.56)$, $Z2=(1.71, 3.08)$ and $Z3=(3.21, 8.26)$. With the blue points we have represented Z1 sample, with the green ones the Z2 sample and with the red points the Z3 sample. The respective fitted lines are in the same colours.}
\label{fig: redshiftbin}
\end{figure*}

The results summarized in Table \ref{Table1} and shown in Fig.\,\ref{fig: redshiftbin} allow us to draw some interesting remarks. Firstly, we note that, although the intrinsic scatter is quite large (mainly because of the use of the fiducial rather than the UP sample), the correlation coefficient $\rho_{LT}$ is quite large in all the redshift bins thus arguing in favour of the existence of LT correlation at any $z$. The slopes $a$ for bins Z1 and Z2 are consistent within the $68\%$ CL, while this is not for bin Z3 where the agreement is present only at the $95\%$ CL level. On the contrary, the zeropoint $b$ is consistent among the three bins. Although a trend in the median values of $a$ is present, we therefore can not conclude that the LT correlation becomes shallower for higher redshift GRBs because of the paucity of the sample and the inclusion of large $u$ GRBs. Larger samples with low $u$ values and a more homogenous redshift sampling are needed to solve this issue.

The study of redshift evolution of the LT correlation is particularly interesting in view of its application to cosmology. If the calibration parameters had significantly changed with $z$, one could have not used the same set of parameters for all the GRBs as, on the contrary, we have usually done. To better clarify this issue, we first remember that the distance modulus

\begin{equation}
\mu = 25 + 5 \log{D_L(z)}
\label{eq: defmu}
\end{equation}
depends on the cosmological parameters as shown by Eq.(\ref{eq: dl}). On the other hand, because of the Eq.(\ref{eq: lx}), the value of $\mu$ for a GRB at redshift $z$ may be inferred by the LT correlation as\,:

\begin{eqnarray}
\mu_{obs}(z) & = & 25 + \frac{5}{2} \log{\left [ \frac{L^*_{X}(T_a)}{4 \pi f_a(T_a, T_p, F_a T_a) (1 + z)^{(- 1 + \beta_a)}}
\right ]} \nonumber \\
~ & = & 25 + \frac{5}{2} \left \{ a \log{ \left [ \frac{T_a}{1 + z}
\right ]} + b \right \} \nonumber \\
~ & - & \frac{5}{2} \log{\left [ 4 \pi f_a(T_a, T_p, F_a T_a)
(1 + z)^{(- 1 + \beta_a)} \right ]}
\label{eq: muobslt}
\end{eqnarray}
with the error estimated as\,:

\begin{equation}
\sigma_{\mu} = \frac{5 \sigma_{D_L}}{D_L(z) \ln{10}} \
\label{eq: sigmamu}
\end{equation}
We stress here that the total uncertainty is obtained by adding up the statistical error from the propagation of the errors on the involved quantities and the intrinsic scatter.

\begin{figure*}[t]
\centering
\includegraphics[width=17cm]{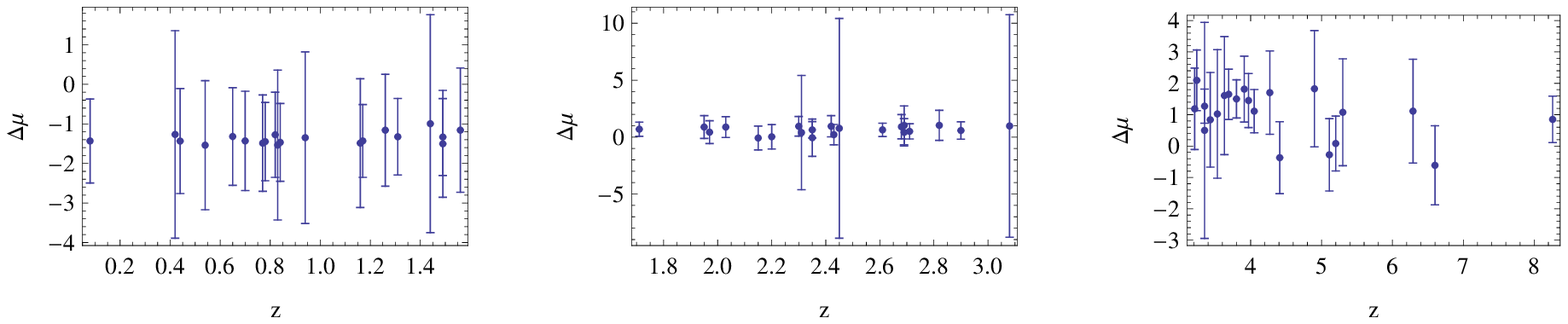}
\caption{$\Delta_{\mu} = \mu_{Zi} - \mu_{fid}$ as function of the redshift $z$ for the three bins Z1 ($0.08 \le z \le 1.56$), Z2 ($1.71 \le z \le 3.08$), Z3 ($3.21 \le z \le 8.26$).}
\label{fig: deltamu}
\end{figure*}

\begin{figure*}[t]
\centering
\includegraphics[width=17cm]{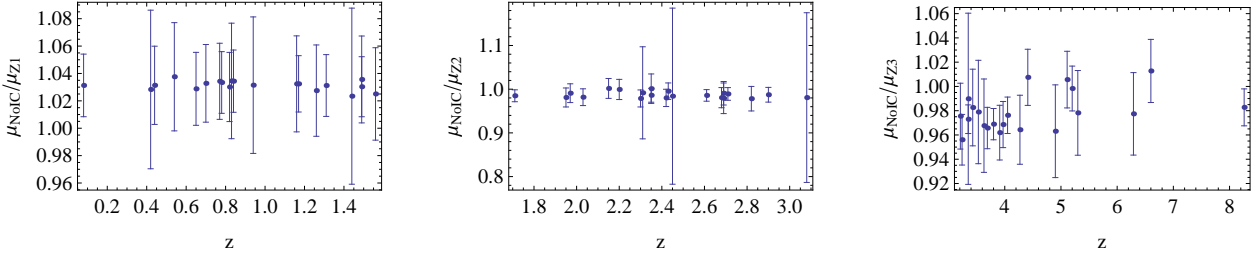}
\caption{Same as Fig.\,\ref{fig: deltamu} but for the distance modulus ratio $\mu_{Zi}/\mu_{fid}$.}
\label{fig: muvsz}
\end{figure*}

We present two tests performed with the aim of understanding if the LT evolves with redshift.
 We denote with $\mu_{Zi}$ and $\mu_{fid}$ the values of $\mu$ estimated from Eq.(\ref{eq: muobslt}) using $(a, b, \sigma_{int})$ obtained by fitting the $Zi$ and the fiducial samples respectively. Fig.\,\ref{fig: deltamu} presents the plots $\Delta_{\mu}= \mu_{Zi} - \mu_{fid}$ vs $z$ for the three redshift bins showing that $\Delta \mu$ is has a roughly constant behaviour in each redshift bin. Moreover, as shown in Fig.\,\ref{fig: muvsz} (giving instead $\mu_{Zi}/\mu_{fid}$ as function of $z$), we can see a flat behaviour too, or at maximum  a difference of order $\sim 2\%$, much smaller than the typical error bars and hence fully negligible. We therefore argue a (still to be confirmed) redshift dependence of the LT correlation does not preclude its usage as a way to construct a GRBs Hubble diagram \citep{Cardone09,Cardone2010} for cosmological applications.

\section{Looking for a redshift estimator}

\begin{figure*}[t]
\centering
\includegraphics[width=14cm]{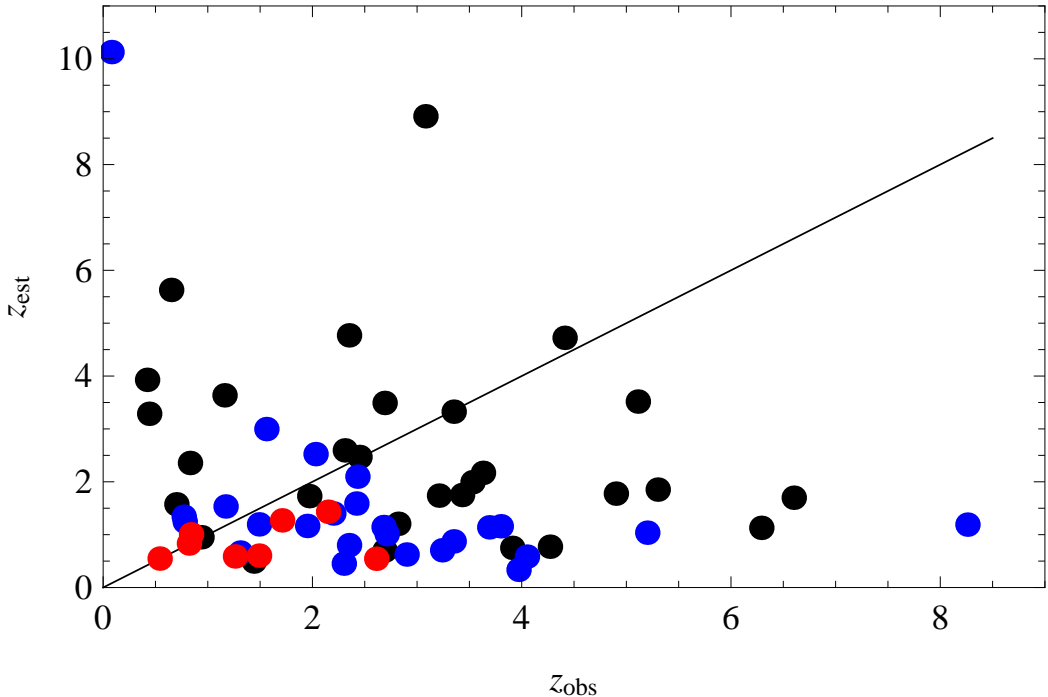}
\caption{Observed vs estimated redshift for the 62 GRBs of the fiducial sample divided in three $u$ bin, i.e, $u \le 0.095$ (red points), $0.095 \le u \le 0.3$ (blue points), $0.3 \le u \le 4$ (black points).}
\label{redshiftestimator}
\end{figure*}

The above analysis has shown that the LT correlation is empirically well motivated and not affected by selection effects due to $u$ selection or to redshift dependent calibration. It is therefore worth investigating its possible applications as redshift estimator. To this aim, let us go back to Eq.(\ref{eq: lx}) and rearrange it in a different way as follows\,:

\begin{eqnarray}
\log{[L^*_X(T_a)]} & = & \log{(4 \pi F_X)} + 2 \log{D_L(z)} - (1 - \beta_a) \log{(1 + z)} \nonumber \\
~ & = & \log{(4 \pi F_X)} + (1 + \beta_a) \log{(1 + z)} + 2 \log{r(z)} + 2 \log{(c/H_0)} \nonumber \\
~ & = & a \log{\left ( \frac{T_a}{1 + z} \right )} + b
\end{eqnarray}
where we have denoted with $r(z)$ the integral in Eq.(\ref{eq: dl}) and, in the last row, we have used the LT correlation with the definition of $T_a^*$. Solving with respect to $z$, we get\,:

\begin{equation}
(1 + \beta_a + a) \log{(1 + z)} + 2 \log{r(z)} = a \log{T_a} + b - \log{(4 \pi F_X)} - 2 \log{(c/H_0)}
\label{eq: zedest}
\end{equation}
For the considered cosmological model, the right hand side of Eq.(\ref{eq: zedest}) depends only on measurable quantities so that one can try solving this equation with respect to $z$ to get an estimate of the GRB redshift. There are, however, some preliminary issues that must be considered. First, both the observable quantities $(T_a, F_X, \beta_a)$ and the LT calibration parameters $(a, b)$ are affected by their own uncertainties. Propagating these errors on the final estimate of $z$ is not analytically possible. Moreover, the uncertainties on $(a, b)$ are not symmetric and the intrinsic scatter $\sigma_{int}$ (also known with its own asymmetric confidence range) adds to the total uncertainty in a nonlinear way.  If we denote by ${\cal{Z}}({\bf p})$ the solution of Eq.(\ref{eq: zedest}) for a given set of parameters ${\bf p} = \{\log{T_a}, \log{F_X}, \beta_a, a, b\})$ and neglect the correlations among the errors, one should estimate the error on $z_{est}$ as\,:

\begin{displaymath}
\sigma = \left [ \sum{\left | \frac{\partial {\cal{Z}}({\bf p})}{\partial p_i} \right |^2 \sigma^2(p_i)} \right ]^{1/2}
\end{displaymath}
where the sum runs over the number of parameters. Actually, such a formula can not be used since, firstly, we do not have an analytical expression for ${\cal{Z}}({\bf p})$ and, secondly, there is actually a non negligible correlation among the parameters (for instance, $b$ is determined from the value of $a$ and $\sigma_{int}$). To fully take into account this issue, for each GRB, we first estimate $z$ setting all the observable quantities $(\log{T_a}, \log{F_X}, \beta_a)$ to their central values and the calibration parameter $(a, b)$ to their best fit values and solve Eq.(\ref{eq: zedest}) to get what we denote as $z_{est}$. \footnote{Regarding the uncertainty we can only provide a rough estimate repeating the procedure described for a large set of randomly generated values of $(a, b)$, derived by interpolating the (normalized) histograms outputted from the Markov chains. We then take the histogram of the $z_{est}$ values thus obtained to find out the $68\%$ confidence range $(z_{min}, z_{max})$ finally defining $\sigma(z_{est}) = [(z_{max} - z_{est}) + (z_{est} - z_{min})]/2$, i.e. we symmetrize the confidence range. It is worth stressing that such an approach implicitly allows us to propagate also the error on $\sigma_{int}$ since the zeropoint $b$ is determined from the values of $(a, \sigma_{int})$ on a case\,-\,by\,-\,case basis. We don't present in the Fig. \ref{redshiftestimator} the errorbars not to cutter the picture and because for the reasons discussed above they can be only an approximated estimate of the real error measurements.}

We have applied the above test\footnote{Remember that, when performing the test, we use the merged chains relative to the fit to the considered sample so that, for instance, the best fit $(a, b)$ values are different in the two cases.} to the fiducial $(u \le 4.0)$ and the U0095 samples $(u \le 0.095)$ finding out that the LT correlation can still not be used as a redshift estimator, as clearly shown in Fig.\,\ref{redshiftestimator}. Indeed, defining $\Delta z = z_{obs} - z_{est}$, we get only $\sim 20\%$ ($28\%$) of GRBs in the fiducial (U0095) sample has $|\Delta z/\sigma(z_{est})| \le 1$. Even if we allow for a very poor precision considering as acceptable estimates those with $|\Delta z/\sigma(z_{est})| \le 3$, the fraction of successful solutions raises only to a modest $\sim 53\%$ ($\sim 57\%$) for the fiducial (U0095) samples. The qualitative agreement about the bad performance of this estimator for both the fiducial and U0095 samples is a first evidence that the value of $u$ has no impact on the quality of the redshift estimate. This is also shown in Fig.\,\ref{redshiftestimator} where the points closer to the $z_{obs} = z_{est}$ line are not the ones with the smaller $u$ values. Actually, such a result can be easily understood noting that, as yet demonstrated above, the $u$ selection does not bias the samples so that the underlying motivation why this redshift estimator fail applies equally to all GRBs notwithstanding how precise is the measurement of $(\log{T_a^{\star}}, \log{L_X^{\star}})$. Actually, the main motivation of the failure is the intrinsic scatter of the points around the best fit LT correlation. It is quite easy to qualitatively understand this point considering a hypothetical GRB of the U0095 sample with given value of $\log{T_a}$. Because of the intrinsic scatter of the LT correlation, its estimated intrinsic luminosity $\log{L_X}$ can be off from the true one up to $\sigma_{int}$. Let us denote with $L_X^{true}$ and $L_X^{fit}$ its true and fitted $L_X$ values and let $L_X(z) = 4 \pi D_L(z)^2 F_X (1 + z)^{-(1 - \beta_a)}$. Our estimate of $z$ is obtained by solving $L_X(z_{est}) = L_X^{fit}$, while it is $L_X(z_{obs}) = L_X^{true}$. Should $L_X^{fit}$ be larger than $L_X^{true}$, we should increase $D_L(z)$ to compensate for the difference thus overestimating the redshift $(z_{est} > z_{obs})$ with the opposite effect for the $L_X^{fit} < L_X^{true}$ case. This is indeed what happens in our case. Looking at the residuals of the LT correlation, we find that $L_X$ is, on average, underestimated (i.e., $L_X^{fit} < L_X^{true}$) for the higher redshift GRBs so that we expect $z_{est} \le z_{obs}$ which is indeed what we find looking at the points with $z > 2.5$ in Fig.\,\ref{redshiftestimator}, while the opposite result takes place for very low $z$ GRBs. We can therefore conclude that the LT correlation works satisfactorily well as redshift estimator only for GRBs in the redshift range $(0.5, 2.0)$, while gives severely biased results for smaller and larger $z$ GRBs.

Although these results are quite discouraging, it is nevertheless worth wondering whether the situation can be improved with future data. To this end, we have simulated a U0095 sample generating $(\log{T_a^{\star}}, \beta_a, z)$ values from a parent distribution closely mimicking the observed one for the fiducial sample\footnote{Note that this choice is motivated by the poor statistics of the present U0095 sample. We have, however, checked that the U0095 GRBs cover the same region in the $(\log{T_a}, \beta_a, z)$ parameter space as the fiducial ones.}. We then set $\log{L_X^{\star}}$ extracting from a Gaussian distribution centered on the value predicted by the LT correlation and with width equal to the intrinsic scatter. We then use these values to estimate $F_X$ and add noise to all quantities so that the relative errors are the same order as the present day ones. We generate ${\cal{N}}$ GRBs and fit them with the same procedure adopted to find the LT calibration coefficients and use these fake Markov chains as input to the redshift estimate procedure.

It turns out that increasing the sample is not a useful way to improve the performance of the redshift estimator. Indeed, we have found that, with ${\cal{N}} \simeq 50$, the fraction of GRBs with $|\Delta z/\sigma(z_{est})| \le 1$ first increases to $\sim 34\%$ and then decreases to $\sim 20\%$ for ${\cal{N}} \simeq 200$, while $\langle \Delta z/z_{obs} \rangle \simeq -17\%$ for both ${\cal{N}} \simeq 50$ and ${\cal{N}} \simeq 200$, a significant improvement with respect to the value quoted above, but still not fully satisfactory considering that $rms(\Delta z/z_{obs}) \simeq 75\%$. It is somewhat counterintuitive that increasing the sample does not improve the quality of the redshift estimator. Actually, such a result could be anticipated noting that a larger sample leads to stronger constraints on the $(a, b, \sigma_{int})$ values, but do not change the intrinsic scatter which is the main source of possible mismatches between the true and fitted GRB luminosity. Motivated by this consideration, we therefore perform a second test artificially lowering the intrinsic scatter $\sigma_{int}$, but setting the best fit $(a, b)$ parameters to those derived from the fit to the U0095 sample. Indeed, for ${\cal{N}} = 50$ and $\sigma_{int} = 0.20$, we get $\langle \Delta z/z_{obs} \rangle \simeq -3\%$ with $rms(\Delta z/z_{obs}) \simeq 28\%$ and $46\%$ ($87\%$) of GRBs with $|\Delta z/\sigma(z_{est}) \le 1 (\le 3)$. Again increasing the sample to ${\cal{N}} \simeq 200$ has not a significant impact, while a stronger impact is obtained setting $\sigma_{int} = 0.10$ giving $\langle \Delta z/z_{obs} \rangle \simeq -0.6\%$ with $rms(\Delta z/z_{obs}) \simeq 16\%$ and $f(|\Delta z/z_{obs}| \le 1) \simeq 66\%$. These results convincingly show that the LT correlation could be used as a redshift estimator only if a subsample of the canonical GRBs could be identified in such a way to reduce the intrinsic scatter to $\sigma_{int} = 0.10 - 0.20$. It is worth wondering whether assembling such a sample is indeed possible. Actually, a detailed answer can not be given since our U0095 sample is too small to find out some indicator that can help to find out GRBs less scattering from the best fit LT correlation. A visual inspection of the fit residuals makes us roughly argue that $\sigma_{int}$ could be reduced using only 5 out of the 8 U0095 GRBs which represent $\sim 8\%$ of the fiducial GRBs sample. If we assume this fraction as a constant, one should assemble a sample of $\sim 600$ GRBs with measured values of $(\log{T_a^{\star}}, \log{L_X^{\star}}, \beta_a, z)$ to get $\sim 50$ GRBs to calibrate the LT correlation with $\sigma_{int} \sim 0.20$. While this is for sure an ambitious task, it is worth noting that it is still possible that a smaller sample is enough to find out the observable properties of these GRBs thus allowing an easier search and reducing the number of GRBs to be followed up for the $z$ estimate.

\section{Summary}

The analysis presented here have shown that the LT correlation, for both the fiducial and UE samples, is not affected by selection effects induced by the $u$ threshold selection or by the implicit assumption of redshift independence of the calibration parameters. In particular, the selection on $u$ does not bias the distribution of the $(\log{L_X^*}, \log{T_a^*}, \beta_a)$ quantities thus showing that the canonical GRBs ($u < 0.095$) in the $U0095$ sample are indeed distributed preferentially in the upper part of the LT plane. This is a further evidence that the afterglow light curves which are smooth and well fitted by the W07 model indeed define a physically homogenous class with the remarkable feature of obeying a well defined empirical correlation. Furthermore, the analysis presented also pinpoints the existence of a well defined correlation of the X\,-\,ray spectral index $\beta_a$ with the rest frame break time $T^*_a$ which deserves further analysis.

As an important result, we have also shown that, although a shallowing of the LT correlation for higher $z$ GRBs can still not be totally excluded, its impact on the distance modulus estimate is negligible thus validating the usage of this correlation as a new independent cosmological tool \citep{Cardone2010}. As a first application, Cardone et al. (2010) have indeed derived the Hubble diagram using the LT correlation only and shown that, when combined with other distance probes (such as Type Ia Supernovae and Baryon Acoustic Oscillations), GRBs are a valuable tool to constrain the cosmological parameters. To this end, Cardone et al. (2010) have used the full fiducial sample to increase the statistics, but at the price of including GRBs with large errors on the distance modulus. It is worth wondering how large a GRB sample should be to improve the constraints on the cosmological parameters. Such a problem has yet been addressed by some of us \citep{Cardone09} using the Fisher matrix analysis and the first version of the LT correlation. There, we have shown that combining a sample of 200 GRBs with a SNAP\,-\,like SNeIa sample allows to determine the matter density parameter $\Omega_M$ and the dark energy equation of state parameters $(w_0, w_a)$ within 0.019, 0.036, 0.020, respectively. In particular, GRBs are of extremely importance to constrain $\Omega_M$ giving an improvement in precision of a factor 4 with respect to the case when SNeIa only are used. Although these results refer to the first version of the LT correlation and thus refer to a sample with no selection on $u$, they qualitatively hold also in our case since the basic inputs to the Fisher matrix analysis are essentially the same. Actually, having made no cut on $u$, the quoted results are likely to be quite conservative since the $u$ selection allows to reduce the scatter and hence the error on the distance modulus thus increasing the efficiency of GRBs with respect to the case considered in Cardone et al. (2009).

We have, finally, investigated the possibility to use the LT correlation as a redshift estimator obtaining actually discouraging results for both the fiducial and U0095 samples. Having qualitatively discovered the reason of this failure, we have shown that reducing the intrinsic scatter of the LT correlation could help to calibrate an improved LT correlation that could work as a tool to estimate the GRB redshift from the analysis of its X\,-\,ray afterglow lightcurve. However, we are not sure if it is possible to obtain a reduced intrisic scatter of the correlation with the real data measurements related to the U0095 sample.

\section*{Acknowledgments}

This work made use of data supplied by the UK Swift Science Data Centre at the University of Leicester. MGD and MO are grateful for the support from Polish MNiSW through the grant N N203 380336. MGD is also grateful for the support from Angelo Della Riccia Foundation.

\end{document}